# An ACO Algorithm for Effective Cluster Head Selection

**Amritha Sampath, Tripti. C, Sabu M. Thampi**
Department of Computer Science and Engineering
Rajagiri School of Engineering and Technology, Kochi, India
amrithasampath@yahoo.com

*Abstract*— **This paper presents an effective algorithm for selecting cluster heads in mobile ad hoc networks using ant colony optimization. A cluster in an ad hoc network consists of a cluster head and cluster members which are at one hop away from the cluster head. The cluster head allocates the resources to its cluster members. Clustering in MANET is done to reduce the communication overhead and thereby increase the network performance. A MANET can have many clusters in it. This paper presents an algorithm which is a combination of the four main clustering schemes- the ID based clustering, connectivity based, probability based and the weighted approach. An Ant colony optimization based approach is used to minimize the number of clusters in MANET. This can also be considered as a minimum dominating set problem in graph theory. The algorithm considers various parameters like the number of nodes, the transmission range etc. Experimental results show that the proposed algorithm is an effective methodology for finding out the minimum number of cluster heads.**

## I. INTRODUCTION

The technology is growing at a faster rate. A day without the use of wireless technology is very rare. It is gaining immense popularity with the use of the most common portable devices like mobile phones, laptops etc. But the need for connectivity in situations where there is no base station available demanded the transition to a new system called 'MANET'- Mobile ad hoc network. These systems are independent consisting of mobile hosts that are connected by multi-hop wireless links. The main feature of this system is that, it lacks a centralised administration or a fixed infrastructure. But implementing the MANET is a difficult task. The absence of a static infrastructure, a boom in a way, increases the challenges to the research community for designing an ad hoc network with issues like resource allocation and routing strategies for the new topology of the ad hoc network [1]. In order to solve these challenges, a clustering mechanism is used to organize the network topology in a hierarchical manner. Many clustering schemes have been developed to solve these issues in MANET. In an ad hoc network, every communication terminal communicates with its neighbor to perform peer-to-peer communication.

MANET consists of group of nodes that can transmit and receive data using wireless links. Two nodes can establish a wireless link among themselves only if the Euclidean distance between them is less than the transmitting range. But communication between nodes which are not in the range of each other is made possible if the other hosts that lie in-between are willing to forward packets for them[2,3]. This characteristic of MANET is known as multi-hopping, which makes them more apt for communication.

Routing in an ad hoc network is quite different from the usual wired networks. Here, the network topology changes rapidly and the conventional routing protocols takes large computational time as well as the result obtained may not satisfy the needs of a dynamic network [4]. The topology of the network may change during the packet transmission also. This situation demands that routes in MANET must be calculated frequently in order to avoid the packet loss [5]. There are different routing schemes in MANET. They can be flooding, proactive routing, reactive routing, and hybrid routing.

Flooding is a distributed process. In flooding, a node transmits a message to all its neighbours and these neighbours will in turn transmit the message to its neighbours. This process continues till the message has been disseminated to the entire network [6]. Flooding is the simplest of all the routing schemes. However, this method is used rarely, because it generates high traffic and as a result, network congestion occurs.

In proactive routing, each node maintains a table containing details regarding its neighbours, the route, distance etc. Whenever a change in the network topology occurs, the valid routes are maintained and updated in all the node tables [2, 7, 8]. This is very efficient for a network with a small size. But, as the network size grows, maintaining the routing table for each node increases network overhead, and thereby it affects the performance of the network.

Reactive routing is also known as on-demand routing. Reactive routing has two main processes involved in it - *route discovery and routing*. If a node needs to find a route to another destination node, first it must discover a route that is needed to reach the destination node and then route the packet to that node. Unlike proactive routing, paths are maintained only until they are needed [9].

Hybrid routing is based on the hierarchical approach [7]. In hybrid routing, the network is organized into a number of small clusters. This reduces the demerits of flooding and pro active routing schemes. Cluster is a subset of nodes. In each cluster, a cluster head is elected. This cluster head is one hop away from all other members of the cluster. Cluster head contains the details of all its cluster members. Consequently, the hierarchical topology decreases the network traffic [2, 10]. Hybrid routing is also known as cluster-based routing. The greatest advantage of this type of routing is that the dynamic topology change appears less dynamic in this scheme [11].

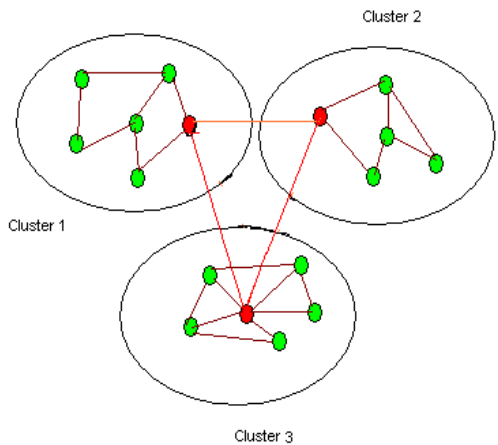

**Figure1: Hierarchical MANET Routing Algorithm**

The MANET can be maintained and managed by partitioning the network into clusters. Clustering is a method used for grouping the nodes based on certain property. The implementation of clustering schemes in MANET helps in improving the routing by reducing the size of the routing table and decreasing transmission overhead by updating the routing tables after topological changes occur[2,10]. It also allows a better performance of the protocols for the MAC layer by improving throughput, scalability and power consumption [12]. The purpose of a clustering algorithm is to produce and maintain a connected cluster [13]. A cluster consists of three types of nodes - *ordinary nodes, gateway nodes and cluster heads.*

Ordinary nodes are cluster members but they do not have neighbors belonging to different clusters [14, 15]. Gateway nodes are nodes in a non-cluster head state located at the boundary of a cluster. They are used for routing to a node from a different cluster. Networks select a set of nodes that can serve as the backbone of the network. A network can contain a number of clusters and each cluster has cluster head and cluster members, which are at one hop away from the cluster head. The clusterhead of one cluster is connected to another cluster directly or through the gateway nodes. The gateway nodes and the cluster heads together manage the routing mechanism of the network [15, 16, 17]. Cluster gateway routing from source node 1 to destination node 8 is shown in the figure 2.

Clusterhead allocates the resources to the other nodes in the network. It must perform extra work with respect to other nodes in the network. The nodes with a high degree of relative scalability are considered for the cluster head selection process [1].

The role of clusterhead is to calculate the routes for messages and to forward inter- cluster packets. A packet from the source node is directed to its clusterhead, if the destination node is also within the same cluster, the clusterhead just forwards the packet to that node. If the destination node is not in the same cluster, the clusterhead of the source node routes the packet to the clusterhead of the destination node and the clusterhead of the destination node then forwards the packet to the destination node [5].

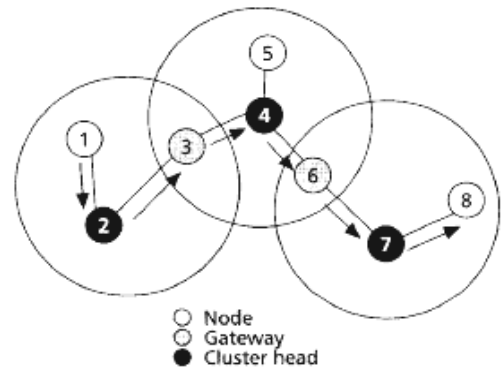

**Figure 2: Cluster Gateway routing from source node 1 to destination node 8 from [28]**

A mobile ad hoc network can be represented as an undirected graph. The network can be represented in the form of a graph as $G = (V, E)$ where V is the set of all nodes and E is the set of all edges- the link between the nodes. There exists a node between two nodes only if both the nodes are in the transmission range R. The distance between two nodes in a graph corresponds to the number of nodes that exist between the nodes. There can be many paths from the source to destination, but the shortest path from the source to destination should be identified for an efficient system [18]. Clustering techniques in an ad hoc network can be based on the dominating set. Many algorithms are proposed to find small dominating sets. A set S is said to be dominating set if and only if each node in the graph $G = (V, E)$ is either in S or adjacent to at least one of the nodes in S [19]. The set of all clusterheads in a network is said to the dominating set and the number of clusterheads in the set is called the domination number [20]. Lesser the domination number, lesser is the communication overhead [21]. Clustering in a mobile ad hoc network can be considered as a minimum dominating set problem. Nevertheless, the computational time for solving this puts the minimum dominating set problem an NP- hard problem.

This paper presents an ACO, a meta- heuristic [22] approach for clustering in MANET. Ant Colony Optimization algorithm uses a colony of artificial ants to find the shortest path between the source node and a destination node by finding the cluster heads in an ad hoc network.

The remainder of this paper is organized as follows: Section 2 describes the state-of-art of clustering algorithms; well-known algorithms are described. In section 3 the proposed ant colony optimization algorithm for effective cluster head selection is discussed. Section 4 presents the studies performed to evaluate the proposed algorithm. Conclusion and topics for further work are presented in the last section.

## II. RELATED WORK

As discussed earlier, a major issue of mobile ad hoc network is its design which is capable of routing dynamically. The routing protocol shall be able to adjust with the high degree of node mobility that often changes the network topology. Static infrastructure is not associated with the mobile ad hoc network. The nodes may have variable amount of resources and this generates a hierarchy in their roles inside the network. Clustering is a hierarchical type of routing in which paths are recorded between clusters instead of between nodes. There are several clustering algorithms based on dominating sets - Lowest ID cluster algorithm (LIC), highest degree algorithm, k-CONID and WCA. A brief overview of each mechanism is discussed below:

### A. Lowest ID cluster algorithm (LIC)

This is an identifier based clustering algorithm in which each node is assigned a distinct ID and the cluster formation is done based on these identifiers. In the lowest ID cluster algorithm, the node with a minimum ID is chosen as the cluster head [23, 24]. A node will always broadcast the list of nodes within its range (including itself). The cluster head is the node that will only hear nodes with ID higher than itself. Thus, the neighbor nodes of the cluster head will be having ID's higher than that of the cluster head. A node that lies within the transmission range of two or more cluster head is called the gateway node and they are used for routing between different clusters in a network.

The drawback of this scheme is that, the lowest ID scheme considers only the lowest node ID that is arbitrarily assigned numbers. It is not considering any other qualifications of a node for selecting the node as a cluster head. The ID of the node does not change with the time and for a long period; the node may have to be the cluster head. Therefore, there is a chance for certain nodes to have power drainage due to serving as clusterheads for longer period of time.

### B. Highest degree algorithm

The highest degree algorithm is also known as connectivity-based algorithm. Here, the degree of a node is calculated based on its distance from the other node. There exists a link between those nodes, if the Euclidean distance between the two nodes is less than the range. In the highest degree algorithm, a node that has maximum degree is chosen as a cluster head [23, 24]. The neighbors of a cluster head become members of that cluster. The algorithm does not limit the number of nodes in a cluster. Therefore, when many nodes are there in the cluster, the throughput drops and the system performance is reduced.

### C. K-hop connectivity ID clustering algorithm (KCONID):

KCONID is a combination of the above said two clustering algorithms: the lowest-ID and highest-degree algorithm [23, 25]. The cluster head is selected based on the connectivity as the initial criterion and then check the lower ID as a secondary criterion. This scheme eliminates the limitations of the above said two algorithms. The purpose of this algorithm is to minimize the number of clusters formed in the network for obtaining the smallest dominating set.

Clusters in KCONID approach are formed by a cluster head and all its nodes are at most distance k- hop from the cluster head. Initially, in the algorithm, a node starts a flooding process in which a clustering request is send to all the nodes. The k-CONID algorithm, generalizes connectivity for a k-hop neighborhood. Thus, when k=1, connectivity is the same as node degree.

The k- CONID works as follows:
1. Each node in the network is assigned a pair of parameters, the connectivity of the node and its identifier.
2. A node is selected as a cluster head if it has the highest connectivity. If two or nodes have the same connectivity, the second criteria - the lowest ID priority is checked to find the cluster head.
3. The idea is that every node broadcasts its clustering decision once all its k- hop neighbors with larger cluster head priority have been done.

### D. Weighted Cluster Algorithm (WCA)

The weighted cluster algorithm elects the cluster head based on the factors like node mobility, number of nodes a cluster head can handle, transmission power etc... [23, 25, 26, 27]. The cluster head must not be over-loaded and therefore a pre-defined threshold value is used which indicates the number of nodes each cluster head can support. The weighted clustering algorithm selects a cluster head according to the weight value of each node. The weight associated to a node $v_i$ is defined as: $W_{vi} = w_1 \Delta_{vi} + w_2 D_{vi} + w_3 M_{vi} + w_4 P_{vi} \ldots\ldots(1)$

The node with the minimum weight is selected as a cluster head. $w_1$, $w_2$, $w_3$, and $w_4$ are weighting factors. The weighting factors are chosen such that $w_1 + w_2 + w_3 + w_4 = 1$. $M_{vi}$ is the measure of mobility. It depends on the average speed of every node during a specified time T. $\Delta_{vi}$ is the degree difference. $D_{vi}$ is defined as the sum of distances from a given node to all its neighbors. This factor is related to the energy consumption since more power is needed for the long distance communication. The parameter $P_{vi}$ is the cumulative time for which a node is retained as the cluster head. This factor is related to measure the power consumption. The cluster head election continues until all the nodes in the network is covered. No two clusterheads can be immediate neighbors.

### III  ACO ALGORITHM - FOR EFFECTIVE CLUSTERHEAD SELECTION

ACO is an evolutionary algorithm. It uses a meta-heuristic approach. Dorigo [22] introduced this nature inspired evolutionary algorithm. It is inspired by the foraging behavior of ants. The ants release chemical called pheromones on the path while moving along the path. As more number of ants moves along the path, the pheromone concentration increases. The more the pheromone concentration, more is the chance for a new ant to choose that path to reach the food from the colony. The path chosen by the ants will be the shortest path from nest to food. The process is a kind of distributed optimization mechanism, in which they find the shortest distance from the food to colony. Every single ant contributes to the solution, cooperating in the work. Artificial ants are used to find the solution of difficult optimization problems. Artificial ants use an incremental constructive approach to search for a feasible solution.

As discussed earlier, the peer-to-peer network can be considered in the form of a graph. In the graph, each node's location is represented using its x and y co-ordinate values and is identified by its unique node number. A node is said to be in the range of another node, if the Euclidian distance between the two nodes is within the range of each other. A hierarchical routing is done using clustering in which paths are recorded between clusters instead of between nodes. This reduces the amount of routing control overhead. ACO finds the minimal set of cluster heads. This is an iterative process and the output obtained is a local solution.

A cluster head is selected based on two aspects, the pheromone value associated with each node and its visibility. Visibility refers to the number of nodes that will be covered if the node is added into the cluster head set [20]. Visibility keeps changing as topology changes. The pheromone value associated with a node is updated for each iteration of the algorithm. For each iteration, a node is selected as the cluster head and the next cluster head is selected based on the pheromone and visibility of its neighbor nodes. This process continues until all the nodes in the network are covered. A node is said to be covered if it is a cluster head or falls in the range of an already selected cluster head.

Each time a node is selected as a cluster head, its pheromone value is updated. Thus, possibility of a node to be selected as cluster head depends on the pheromone value and visibility which changes as the algorithm proceeds through the various iterations. The probability of each node to be selected as cluster head is calculated based on these two values.

$$P_{vi}(iter) = \frac{Wt_{vi}*\alpha + [ph_{vi}(iter)]*\beta}{\sum_{vi=0}^{n} Wt_{vi}*\alpha + [ph_{vi}(iter)]*\beta} \quad ....(2)$$

The equation (2) gives the probability of each node to be selected as a cluster head. It is represented by $P_{vi}$. $Wt_{vi}$ is the weight associated with each node $v_i$. $Wt_{vi}$ is dependent on the degree associated with each node.

$$Wt_{vi} = Degree_{vi} + 1........(3)$$

$ph_{vi}(iter)$ is the pheromone concentration associated with each node for an iteration. $\alpha$ controls the relative importance of visibility measure and $\beta$ controls the pheromone value.

The process can be represented as follows:

```
Initialize each node
     weight=0
     isHead=false
     pheromone=0
     convector=empty  //To store al neighbor
     nodes According to instantaneous topology,
     Find neighbor   //nodes in Range
          If Euclidian distance < Range
               Edge exists
          Else
               No edge exists
     Increment weight // connectivity measure
Set α, β values           // control parameters
Iterate i to n times
     Select i^th node as Cluster head
     Select neighbor with maximum probability
     as next cluster head until all nodes are
     covered
     Update pheromone of the selected cluster
     heads
Find final set of cluster heads with maximum
probability (with Weight and Updated pheromone)
```
**Fig. 2 Procedure for clusterhead selection**

The algorithm shows that the $P_{vi}$, the probability of an ant to choose a node is proportional to the degree of each node and the pheromone concentration factor.

### IV  EXPERIMENTAL SIMULATION

*A. Simulation Setup*

The proposed solution was implemented on an ad-hoc peer-to-peer network by placing a specific number of nodes in an MxM area. Nodes were placed randomly and the proposed ACO algorithm was performed on three networks with varying number of nodes and transmission range.

The ACO algorithm was simulated in a Java-eclipse platform. Three networks were generated with sizes 50, 100, 200, 300 and 400 respectively. The number of clusters formed and the other parameters are shown in table 2. The algorithm was iterated number of times equal to the number of nodes for an optimal solution. The smallest dominating set for each graph was recorded. Once a node is selected as the cluster head, all the nodes, which is one hop away from the cluster head will be covered and thus forms the cluster. The model of the ad hoc network developed for 200 nodes is shown in figure 3. The nodes are shown as circles with an identifier associated with each. Table 1 shows the ACO parameters used in this work.

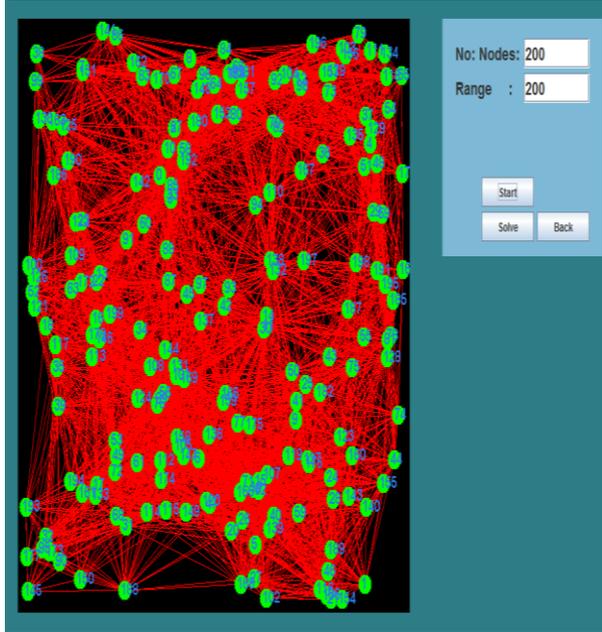

**Figure 3: Mobile Ad hoc network with 200 nodes. The range is arbitrarily chosen as 200.**

**Table 1. The ACO parameters and their values**

| ACO parameters used in the study | |
|---|---|
| α | 9 |
| β | 1 |
| Number of ants | 20 |

### B. Simulation Result Analysis

The proposed solution was tested for five networks of sizes 50, 100, 200, 300 and 400 nodes. Table 2 shows the results of the experiments using the parameters in Table1. The cluster formation of a network with 200 nodes is shown here in figure 5.

Results shows that the as the range increases, the number of clusters formed decreases, i.e. as range of a mobile node increases, connectivity or the weight of a node also increases. So, selection of a node as cluster head causes more number of nodes to get covered. For this reason, lesser number of nodes are enough to cover the entire network.

The proposed solution is a combination of both weighted clustering approach and probability based clustering approach used with ant colony optimization algorithm. This scheme can produce results close to the optimal result. The complexity of this algorithm is $O(n^2)$.

**Table 2. The results of running the ACO algorithm on three ad-hoc networks with various transmission ranges**

| No. | No. of nodes | Range | No. of iterations | No. of clusters |
|---|---|---|---|---|
| 1 | 50 | 200 | 25 | 7 |
| | | 300 | 25 | 4 |
| | | 400 | 25 | 2 |
| 2 | 100 | 200 | 50 | 8 |
| | | 300 | 50 | 4 |
| | | 400 | 50 | 2 |
| 3 | 200 | 200 | 100 | 8 |
| | | 300 | 100 | 4 |
| | | 400 | 100 | 2 |
| 4 | 300 | 200 | 150 | 8 |
| | | 300 | 150 | 4 |
| | | 400 | 150 | 2 |
| 5 | 400 | 200 | 200 | 9 |
| | | 300 | 200 | 4 |
| | | 400 | 200 | 2 |

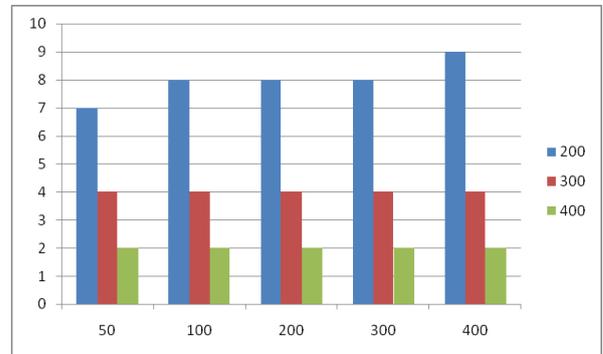

**Figure 4: Comparison of the three networks with different size. The first bar shows the number of clusters when range is 200, second for 300 and third for 300**

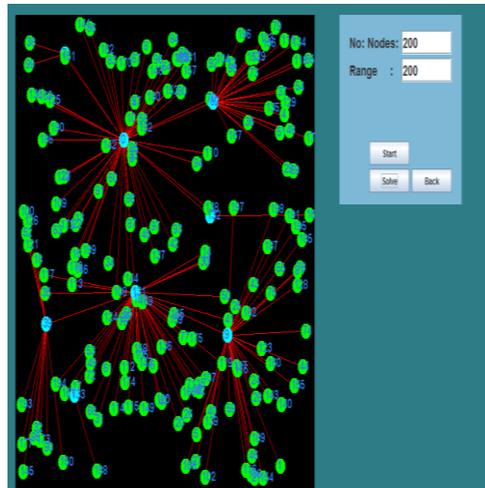

**Figure 5: An ad hoc network with 200 nodes and out of them seven are clusterheads. The clusterheads are denoted by the node in blue colour**

## V. CONCLUSIONS AND FUTURE WORK

The work proposed the use of the ant colony optimization algorithm - a meta-heuristic approach to select the smallest number of clusterheads in an ad hoc network. This is equivalent to finding the minimum dominating set for the topology graph, where each cluster head is a member of the dominating set. Clusterhead is one hop away from its cluster members. The proposed ACO algorithm succeeded in finding out the minimum number of cluster heads with 'n' number of iterations, where 'n' is the number of nodes in the network. In future, the proposed system combined with soft computing techniques such as genetic algorithm, neural network or fuzzy logic may offer a solution more close to global optima.